\documentstyle[aps,prl,epsfig]{revtex}
\begin{document}
\title{Parity Violating Bosonic Loops at Finite Temperature}
\author{V. S. Alves$^{a,b}$, Ashok Das$^{b}$, Gerald V. Dunne$^{c}$
and Silvana  Perez$^{a,b}$}
\address{$^{a}$ Departamento de F\'{\i}sica, Universidade Federal do
Par\'{a},  66075-110 Bel\'{e}m, Brasil}
\address{$^{b}$ Department of Physics and Astronomy,
University of Rochester, Rochester, NY 14627-0171, USA}
\address{$^{c}$ Department of Physics, University of Connecticut, Storrs, CT
06269-3046, USA}

\maketitle
\vskip .5cm

\begin{abstract} 
The finite temperature parity-violating contributions to the polarization tensor
are computed at one loop in a system without fermions. The system studied is a
Maxwell-Chern-Simons-Higgs system in the broken phase, for which the
parity-violating terms are well known at zero temperature. At nonzero
temperature the static and long-wavelength limits of the parity
violating terms have very different structure, and involve non-analytic log
terms depending on the various mass scales. At high temperature the boson loop
contribution to the Chern-Simons term goes like T in the static limit and like
$T\log T$ in the long-wavelength limit, in contrast to the fermion loop
contribution which behaves like $1/T$ in the static limit and like
$\log T/T$ in the long wavelength limit.

\end{abstract}

\vskip 1cm

\section{Introduction}
The study of induced Chern-Simons terms in $2+1$ dimensional field theory at
finite temperature has produced some interesting new insights into large gauge
invariance and parity-violating effective actions at finite temperature
\cite{dunne1,deser1,schaposnik1,lhl,bdf}. Almost all previous studies (with
the exception of \cite{pisarski1}) have concentrated on the induced Chern-Simons
term arising from fermion loop contributions to the gauge field self-energy in
$2+1$ dimensions. In this paper we study finite temperature induced
Chern-Simons terms in a bosonic theory. The induced Chern-Simons terms are
generated in radiative loop corrections, due to the presence of a bare
Chern-Simons term. Specifically, we consider a Maxwell-Chern-Simons-Higgs model
in the spontaneously broken phase. At zero temperature, induced Chern-Simons
terms in such models have been studied in great detail, revealing an intricate
relation between spontaneous parity violation and spontaneous symmetry breaking
\cite{khlebnikov,int,khare,manu}. One motivation for this present paper is to
understand how this generalizes to finite temperature.

The induced Chern-Simons coefficient is extracted from the zero-momentum limit of
the parity violating part of this self-energy \cite{redlich}. At finite T, this
procedure is not unique \cite{kao1} since Feynman diagrams are not analytic in
external momenta at finite temperature \cite{weldon}, because the thermal heat
bath breaks Lorentz invariance. In a static limit, with $q^0=0$ and $|\vec{q}|\to
0$, an induced Chern-Simons term is found with a temperature dependent
coefficient \cite{babu}. As first pointed out in \cite{pisarski1}, this result
appears (when carried over to a non-Abelian theory) to violate large gauge
invariance since the coefficient of the induced Chern-Simons term in a
non-Abelian theory should take discrete values \cite{deser2}. This puzzle has
been resolved for the fermion loop when the background has the character of a
static Abelian magnetic field with integer flux $\Phi$, because in this case the
problem factorizes into $\Phi$ copies of an exactly solvable $0+1$-dimensional
model \cite{dunne1,deser1,schaposnik1,lhl}. Then one finds that the finite
temperature effective action has an infinite series of parity-violating
terms (of which the Chern-Simons term is only the first), each of which
has a T dependent coefficient at finite T. Nevertheless, the series is
such that the full effective action changes under a large gauge
transformation in a manner that is independent of T. These new
parity-violating terms are non-extensive (i.e., they are not integrals of
a density) and they explicitly vanish at zero temperature (as they must
since the zero T effective action should be extensive). This issue of large
gauge invariance of the finite temperature effective action is considerably more
difficult to resolve in genuinely time-dependent and genuinely non-Abelian
backgrounds, although much recent progress has been made in understanding the
parity-violating parts of multi-leg amplitudes at finite temperature \cite{bdf}.

Another motivation for our study is the question of the analytic structure of the
bosonic self-energy at finite temperature. This issue has been analyzed
previously \cite{arnold} for massive gauge bosons in four dimensional space-time,
where the Chern-Simons parity-violating issues are not relevant. In the four
dimensional case it was found that in the broken phase the different bosonic
masses appearing in the bosonic loop meant that the zero energy-momentum limit
was actually analytic, despite the well-known physical difference between the
Debye and plasmon masses, which can be defined through the static and long
wavelength limits, respectively \cite{arnold}. In this current paper, we find
that in three dimensional space-time, for a model with parity violation, the zero
energy-momentum limit is not unique, even though the bosonic masses entering the
one-loop calculation are different.

In Section II we define the bosonic model to be studied, and present the finite
temperature propagators necessary for a perturbative analysis. In Section III
we present the one loop results for the parity violating part of the finite
temperature self energy in both the static and long wavelength limits. Section
IV contains our conclusions.

\section{Maxwell-Chern-Simons-Higgs Model}

We consider an Abelian gauge field $A_\mu$ in $2+1$ dimensions with both
a Maxwell and a Chern-Simons term in the Lagrangian, interacting with a
charged scalar field $\Phi$ which has a symmetry breaking quartic
potential:
\begin{eqnarray} {\cal L}=-\frac{1}{4}F_{\mu\nu}F^{\mu\nu}+
\frac{\kappa}{2}\epsilon^{\mu\nu\lambda}A_\mu\partial_\nu A_\lambda
+|D_\mu \Phi |^2-\frac{\lambda}{4}\left(|\Phi|^2-v^2\right)^2
\label{mcshlag}
\end{eqnarray} In the spontaneously broken phase, where $\Phi$ has a
nonzero vacuum expectation value $\langle \Phi\rangle=v$, we expand the
scalar field about this v.e.v. as
$\Phi=v+\frac{1}{\sqrt{2}}(\sigma+i\chi)$ and obtain the following
Lagrangian in the $R_{\xi}$ gauge (ignoring the ghost Lagrangian which
is not relevant to our calculations):

\begin{eqnarray} {\cal L} &=& -\frac{1}{4} F_{\mu \nu} F^{\mu \nu} + 
\frac{\kappa}{2}\epsilon^{\mu \nu \lambda}
A_{\mu}\partial_{\nu}A_{\lambda} +
\frac{m^2}{2}A_{\mu}A^{\mu} -
\frac{1}{2\xi}\left(\partial_{\mu}A^{\mu}\right)^2+
\frac{1}{2}\partial_{\mu}\sigma \partial^{\mu} \sigma - 
\frac{1}{2}m_{\sigma}^2 \sigma^2 + \frac{1}{2} \partial_{\mu} \chi
\partial^{\mu} \chi -\frac{1}{2} m_{\chi}^2\chi^2\nonumber\\  &-& e
\sigma\stackrel{\leftrightarrow}{\partial^{\mu}} \chi A_{\mu} +
\frac{e^2}{2}(\sigma^2 + \chi^2 + 2 \sqrt{2} v \sigma)
A_{\mu}A^{\mu}-
\frac{\lambda}{2 \sqrt{2}} v \sigma(\sigma^2 + \chi^2) -
\frac{\lambda}{16}(\sigma^2+\chi^2)^2
\label{fluclag}
\end{eqnarray} 
Here the various mass parameters are
\begin{eqnarray} m^2&=&2 e^2 v^2\nonumber\\ m_{\sigma}^2&=&\lambda
v^2\nonumber\\ m_{\chi}^2 &=&  \xi m^2
\end{eqnarray} 
As mentioned above in the Introduction, for the corresponding system {\it
without} the Chern-Simons term (i.e. for $\kappa=0$), the finite temperature
behavior of the polarization tensor was studied in \cite{arnold}. There, one of
the key features was the difference between the bosonic masses appearing in the
one-loop calculation. The model {\it with} a Chern-Simons term is more
interesting for two reasons. First, the presence of the Chern-Simons term leads
to a different mass generation mechanism for the gauge field, with it
acquiring two (rather than one) massive modes in the broken phase \cite{pr}.
Second, the Chern-Simons coupling leads to parity-violating contributions
to the polarization tensor, whose finite temperature behavior is the
subject of this paper. Both these differences can be seen clearly in the
propagator structure of the model. 

\subsection{Zero Temperature Propagators} 
At zero temperature, the gauge
field propagator is
\begin{eqnarray} 
D_{\mu \nu} (p) =\frac{-1}{(p^2 - m_+^2 + i\epsilon) (p^2 -
m_-^2 + i\epsilon)}\,\left[ \eta_{\mu \nu}(p^2 - m^2) -
 p_{\mu} p_{\nu} \frac{(1-
\xi)(p^2 - m^2) + \xi \kappa^2 }{p^2 - \xi m^2} + i \kappa
\epsilon_{\mu
\nu\lambda} p^{\lambda}\right]
\label{ztprop}
\end{eqnarray} where the two massive modes are identified by the poles at
\begin{eqnarray} m_{\pm}^2 = \frac{\kappa^2 + 2 m^2 \pm |\kappa| (\kappa^2
+ 4 m^2)^{1/2}}{2}
\label{mspm}
\end{eqnarray} from which we deduce the (positive) masses
\begin{eqnarray}
m_\pm=\frac{|\kappa|}{2}\left(\sqrt{1+\frac{4m^2}{\kappa^2}}\pm 1\right)
\label{mpm}
\end{eqnarray} 
Note also the presence in $D_{\mu\nu}(p)$ of the term proportional to
$\epsilon_{\mu\nu\lambda} p^{\lambda}$, which manifestly breaks
parity.  The scalar
field $\sigma$ has the standard bosonic propagator $D_{\sigma} (p) =
1/(p^2 - m_{\sigma}^2)$.

\subsection{Finite Temperature Propagators}

At finite T, propagators can be presented either in the imaginary-time or
real-time formalism \cite{kapusta,lebellac,dasbook}. Here we record the
propagators in both forms for the model in (\ref{fluclag}).

\subsubsection{Imaginary time:}

In the imaginary-time formalism, the gauge field propagator is
($\kappa\rightarrow i\kappa$ in the Euclidean space)

\begin{eqnarray} D_{\mu \nu}^{(\beta)} (p) &=&
\frac{1}{(w_n^2 + \vec{p}^2 + m_+^2) (w_n^2
+\vec{p}^2+m_-^2)} \left[ \delta_{\mu
\nu}(w_n^2+\vec{p}^2+ m^2) - 
 p_{\mu} p_{\nu} \frac{(1- \xi)(w_n^2+\vec{p}^2+ m^2) - \xi \kappa^2
}{w_n^2+\vec{p}^2+\xi m^2} -  \kappa \epsilon_{\mu \nu \lambda}
p_{\lambda}\right]
\end{eqnarray} and the scalar $\sigma$ field propagator is
\begin{eqnarray} D_{\sigma}^{(\beta)} (p) = \frac{1}{w_n^2+\vec{p}^2+
m_{\sigma}^2}
\end{eqnarray} Here, $-ip^{0}\rightarrow \omega_n=2\pi n T$ are the
usual  bosonic Matsubara modes.

\subsubsection{Real time:}

In the real-time formalism, the degrees of freedom are doubled in the
standard way \cite{dasbook} in order to account for the transfer of
energy into and out of the thermal heat bath. The propagators thus
acquire a $2\times 2$ matrix structure, the components of which are
listed below, in the closed time path formalism, for the MCSH system
in the  broken phase. For the gauge field,

\begin{eqnarray*} D_{\mu \nu}^{(\beta)++} (p) &=& -\left[ \eta_{\mu
\nu}(p^2 - m^2) - p_{\mu} p_{\nu} \frac{(1- \xi)(p^2 - m^2) + \xi \kappa^2
}{p^2 - \xi m^2} + i  \kappa \epsilon_{\mu \nu \lambda}
p^{\lambda}\right]\\ &\times& \left[\frac{1}{(p^2 - m_+^2 +i 
\epsilon) (p^2 - m_-^2 + i  \epsilon)} - 2 i  \pi n_B(|p^0|)
\delta ((p^2-m_+^2)(p^2-m_-^2))\right]
\end{eqnarray*}

\begin{eqnarray*} D_{\mu \nu}^{(\beta)+-} (p) &=& 2 i  \pi \left[
\eta_{\mu \nu}(p^2 - m^2) - p_{\mu} p_{\nu} \frac{(1- \xi)(p^2 - m^2) +
\xi \kappa^2 }{p^2 - \xi m^2} + i  \kappa \epsilon_{\mu \nu \lambda}
p^{\lambda}\right]\\ &\times& \left[\theta(-p^0) +
n_B(|p^0|)\right]\delta((p^2-m_+^2)(p^2-m_-^2))
\end{eqnarray*}

\begin{eqnarray*} D_{\mu \nu}^{(\beta)-+} (p) &=& 2 i  \pi \left[
\eta_{\mu \nu}(p^2 - m^2) - p_{\mu} p_{\nu} \frac{(1- \xi)(p^2 - m^2) +
\xi \kappa^2 }{p^2 - \xi m^2} + i  \kappa \epsilon_{\mu \nu \lambda}
p^{\lambda}\right]\\ &\times& \left[\theta(p^0) +
n_B(|p^0|)\right]\delta((p^2-m_+^2)(p^2-m_-^2))
\end{eqnarray*}

\begin{eqnarray*} D_{\mu \nu}^{(\beta)--} (p) &=& -\left[ \eta_{\mu
\nu}(p^2 - m^2) - p_{\mu} p_{\nu} \frac{(1- \xi)(p^2 - m^2) + \xi \kappa^2
}{p^2 - \xi m^2} + i  \kappa \epsilon_{\mu \nu \lambda}
p^{\lambda}\right]\\ &\times& \left[\frac{-1}{(p^2 - m_+^2 -i 
\epsilon) (p^2 - m_-^2 - i  \epsilon)} - 2 i  \pi n_B(|p^0|)
\delta ((p^2-m_+^2)(p^2-m_-^2))\right]
\end{eqnarray*} For the scalar $\sigma$ field,

\[ D_{\sigma}^{(\beta)++} (p) = \frac{1}{p^2 - m_{\sigma}^2 + i 
\epsilon} - 2 i  \pi n_B(|p^0|) \delta(p^2-m_{\sigma}^2)
\]

\[ D_{\sigma}^{(\beta)+-} (p) = -2 i  \pi
\left[\theta(-p^0)+n_B(|p^0|)\right] \delta(p^2-m_{\sigma}^2)
\]

\[ D_{\sigma}^{(\beta)-+} (p) = -2 i  \pi
\left[\theta(p^0)+n_B(|p^0|)\right] \delta(p^2-m_{\sigma}^2)
\]

\[ D_{\sigma}^{(\beta)--} (p) = \frac{-1}{p^2 - m_{\sigma}^2 - i 
\epsilon} - 2 i  \pi n_B(|p^0|) \delta(p^2-m_{\sigma}^2)
\]

\section{One-loop Results}

In this Section we compute the parity-violating part,
$\Pi_{\mu\nu}^{(PV)}$, of the polarization tensor $\Pi_{\mu\nu}$, as
represented by the one-loop Feynman diagram in Fig. 1. The
parity-violating contribution arises from the
$\epsilon_{\mu\nu\lambda} k^{\lambda}$ part of the gauge propagator. We
first review briefly the zero temperature result.

\begin{figure}[hbt]
\centering{\epsfig{file=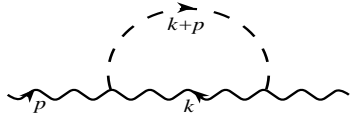, width=12cm, height=7cm}}
\vskip -2cm
\caption{One loop self-energy diagram for computing the induced parity violating
Chern-Simons coefficient. The wavy line represents the gauge field and the
dashed line represents the scalar particle.
\label{f1}}
\end{figure}

\subsection{Zero temperature parity-violating part}

The parity-violating part of the diagram in Fig. 1 is
\begin{eqnarray}
\Pi^{\mu\nu (PV)}= 8i\kappa e^4v^2\epsilon^{\mu\nu\lambda}\int
\frac{d^3k}{(2\pi)^3}\,
\frac{1}{((k+p)^2-m_\sigma^2)}\, \frac{k_\lambda}{(k^2-m_+^2)(k^2-m_-^2)}
\end{eqnarray} A straightforward use of Feynman parameters shows that
this can be expressed as
\begin{eqnarray}
\Pi^{\mu\nu (PV)}=
- \epsilon^{\mu\nu\lambda}p_{\lambda} \Pi(p^2)
\end{eqnarray} 
where
\begin{eqnarray}
\Pi(p^2)=16i\kappa e^4 v^2\, \int
\frac{d^3k}{(2\pi)^3}\, \int_0^1d\alpha \int_0^{1-\alpha}d\beta\,
\frac{\alpha}{(k^2+\alpha(1-\alpha)p^2-\alpha m_\sigma^2-\beta
m_+^2-(1-\alpha-\beta)m_-^2)^3}
\end{eqnarray} 
The induced Chern-Simons coefficient is deduced from the
value of $\Pi(p^2=0)$:
\begin{eqnarray}
\Pi(p^2=0)&=&16i\kappa e^4 v^2 \int_0^1d\alpha \int_0^{1-\alpha}d\beta\,
\frac{1}{32\pi}\,\frac{\alpha}{(\alpha(m_\sigma^2-m_-^2)+\beta(
m_+^2-m_-^2)+m_-^2)^{3/2}}\nonumber\\ &=&\frac{-2i\kappa e^4v^2}{3\pi
(m_\sigma^2-m_+^2)^2 (m_\sigma^2-m_{-}^2)^2
(m_+^2-m_-^2)}\left\{-m_{\sigma}^5(m_+^2-m_-^2) + 2
m_{\sigma}^4(m_+^3-m_-^3)\right.\nonumber\\ &&\left. -
m_{\sigma}^3(m_+^4-m_-^4)-4 m_{\sigma}^2(m_+^3m_-^2 - m_+^2m_-^3)
+ 3 m_{\sigma}(m_+^4m_-^2 - m_+^2m_-^4) -  2
(m_+^4m_-^3 -m_+^3m_-^4)\right\}
\label{ztpi}
\end{eqnarray} 
Notice that the dependence on the three different masses,
$m_\sigma$, $m_+$ and $m_-$, is quite involved.

At this stage we pause to compare with some previous results
corresponding to special cases of this result. In the pure Chern-Simons
limit, in which the Maxwell term is removed from the Lagrangian, the
corresponding result was computed in \cite{khlebnikov}. This limit can be
obtained from our result by sending $e^2\to\infty$ and
$\kappa\to\infty$, in such a way that the ratio $\frac{e^2}{\kappa}$ is
kept finite. In terms of the masses, in this limit $m_+\to\infty$,
$m_-\to\frac{m^2}{|\kappa|}=\frac{2e^2v^2}{|\kappa|}$ (finite), and
$m_\sigma$ is unaffected. In this limit, our result reduces to
\begin{eqnarray}
\Pi(p^2=0)=\frac{2ie^2}{3\pi}\,\frac{\kappa}{|\kappa|}\,
\frac{\left(1+\frac{1}{2}\frac{m_\sigma}{m_-}\right)}
{\left(1+\frac{m_\sigma}{m_-}\right)^2}
\label{ztpcs}
\end{eqnarray}
which is in agreement with \cite{khlebnikov}. Furthermore, when the remaining
masses, $m_\sigma$ and $m_-$, are equal, this gives
\begin{eqnarray}
\Pi(p^2=0)=\frac{ie^2}{4\pi}\,\frac{\kappa}{|\kappa|}
\label{ztsusy}
\end{eqnarray} 
This is exactly the mass relationship ($m_+\to\infty$ and $m_\sigma=m_-$) that
arises in the N=2 supersymmetric Chern-Simons-Higgs system \cite{susycs}, and
this result (\ref{ztsusy}) agrees with the known result for this SUSY model
\cite{lee}. The fact that (\ref{ztsusy}) is equal in magnitude, but opposite in
sign, to the fermion-loop contribution \cite{redlich} to the parity-violating
part of the polarization tensor, reflects the non-renormalization of the
Chern-Simons coefficient in the N=2 SUSY model.

\subsection{Finite temperature parity-violating part}

We now consider the calculation of the finite temperature
one-loop parity-violating part of the polarization tensor. Such a calculation
can be performed either using the imaginary time or the real time formalism of
finite temperature field theory. In this paper we record the imaginary time
calculation; we have also done the calculation using the real time
formalism (the appropriate amplitudes to compare are the retarded ones), and
obtain exactly the same results.

In the imaginary-time formalism, the parity-violating part of
$\Pi_{\mu\nu}$ is
\begin{eqnarray}
\Pi_{\mu\nu (PV)}=  8\kappa e^4 v^2
\epsilon_{\mu\nu\lambda}\frac{1}{\beta}\sum_{n=-\infty}^\infty
\int\frac{d^2k}{(2\pi)^2}\,
\frac{k_\lambda}{((k^0+p^0)^2+(\vec{k}+\vec{p})^2+m_\sigma^2)
(k_0^2+\vec{k}^2+m_+^2)(k_0^2+\vec{k}^2+m_-^2)}
\end{eqnarray} 
where the Matsubara energies are $k^0=\frac{2\pi n}{\beta}$ and 
$p^0=\frac{2\pi l}{\beta}$, with $n$ and $l$ being integers.

\subsubsection{Static Limit}

At finite temperature there are different physical limits for the external
energy-momentum $p$, due to the preferred Lorentz frame of the heat bath. These
different limits reflect  different physical processes \cite{dasbook}. We first
consider the {\it static limit} in which we first set $p^0=0$, and then take the
limit $|\vec{p}|\to 0$. First, observe that in this static limit,
\begin{eqnarray}
\Pi^{{\rm static}\,(PV)}_{ij}=0
\end{eqnarray} since the $k^0$ sum (i.e. the sum over $n$) clearly
vanishes when 
the index $\lambda=0$. The remaining parity-violating components are:
\begin{eqnarray}
\Pi^{{\rm static}\,(PV)}_{0i}= 8\kappa e^4v^2
\epsilon_{ij}\frac{1}{\beta}\sum_{n=-\infty}^\infty
\int\frac{d^2k}{(2\pi)^2}\,
\frac{k_j}{((k^0)^2+(\vec{k}+\vec{p})^2+m_\sigma^2)
(k_0^2+\vec{k}^2+m_+^2)(k_0^2+\vec{k}^2+m_-^2)}
\label{static0i}
\end{eqnarray} The induced Chern-Simons coefficient is determined by the
coefficient of
$\epsilon_{ij}p_j$ in the limit $|\vec{p}|\to 0$, so we look for the
term linear in the spatial momentum $\vec{p}$. Thus, we expand
\begin{eqnarray}
\frac{1}{((k^0)^2+(\vec{k}+\vec{p})^2+m_\sigma^2)}=
\frac{1}{((k^0)^2+\vec{k}^2+m_\sigma^2)}-
\frac{2\vec{p}\cdot\vec{k}}{((k^0)^2+\vec{k}^2+m_\sigma^2)^2}+O(\vec{p}^2)
\end{eqnarray} The first term in this expansion contributes 0 when the
spatial $\vec{k}$ momentum integral is done in (\ref{static0i}). However,
the second term produces a term linear in $\vec{p}$. Using symmetric
integration, we replace
$k_ik_j\to\frac{1}{2}\vec{k}^2\delta_{ij}$, 
and obtain
\begin{eqnarray}
\Pi^{{\rm static}\,(PV)}_{0i}=\epsilon_{ij}p_j\,\Pi_{\rm
static}(\vec{p}^2)
\end{eqnarray} 
where
\begin{eqnarray}
\Pi_{\rm static}(\vec{p}^2=0)&=& - 8\kappa e^4v^2 \frac{1}{\beta}
\sum_{n=-\infty}^\infty \int\frac{d^2k}{(2\pi)^2}\,
\frac{\vec{k}^2}{((k^0)^2+\vec{k}^2+m_\sigma^2)^2
(k_0^2+\vec{k}^2+m_+^2)(k_0^2+\vec{k}^2+m_-^2)}\nonumber\\ &=& 8\kappa
e^4v^2 \frac{\partial}{\partial m_\sigma^2}\, \frac{1}{\beta}
\sum_{n=-\infty}^\infty \int\frac{d^2k}{(2\pi)^2}\,
\frac{\vec{k}^2}{((k^0)^2+\vec{k}^2+m_\sigma^2)
(k_0^2+\vec{k}^2+m_+^2)(k_0^2+\vec{k}^2+m_-^2)}
\label{stat}
\end{eqnarray}

It is convenient to perform the sum over Matsubara modes using the
Sommerfeld-Watson transformation \cite{mw,dasbook} of the sum into a
contour integral:
\begin{eqnarray}
\sum_{n=-\infty}^\infty f(n)=-\pi \sum_{\rm residues}\left(f(z) \cot(\pi
z)\right)
\end{eqnarray} 
where the sum is over the residues at the poles of $f(z)$.
Thus, defining
\begin{eqnarray}
\omega_\sigma=\sqrt{\vec{k}^2+m_\sigma^2}\quad, \quad
\omega_+=\sqrt{\vec{k}^2+m_+^2}\quad, \quad
\omega_-=\sqrt{\vec{k}^2+m_-^2}
\label{omegas}
\end{eqnarray} 
we find that
\begin{eqnarray}
&&\frac{1}{\beta}\sum_{n=-\infty}^\infty \, \frac{1}{\left((\frac{2\pi
n}{\beta})^2+\omega_\sigma^2\right)\left((\frac{2\pi
n}{\beta})^2+\omega_+^2\right)\left((\frac{2\pi
n}{\beta})^2+\omega_-^2\right)}\nonumber\\ &=&
\frac{1}{2}\left[\frac{\frac{1}{\omega_\sigma}\,{\rm coth}(\frac{\beta
\omega_\sigma}{2})}{(m_\sigma^2-m_+^2)(m_\sigma^2-m_-^2)}
-\frac{\frac{1}{\omega_+}\,{\rm coth}(\frac{\beta
\omega_+}{2})}{(m_\sigma^2-m_+^2)(m_+^2-m_-^2)}
+\frac{\frac{1}{\omega_-}\,{\rm coth}(\frac{\beta
\omega_-}{2})}{(m_\sigma^2-m_-^2)(m_+^2-m_-^2)}\right]
\label{sw}
\end{eqnarray}
We can separate the zero temperature contribution from the finite
temperature correction by using the simple identity
\begin{eqnarray} 
{\rm coth}(\frac{x}{2})=1+\frac{2}{e^x-1}
\label{identity}
\end{eqnarray} 
in which we recognize the Bose-Einstein distribution
function $n(x)=\frac{1}{e^x -1}$. Then the zero temperature
contribution can be expressed as
\begin{eqnarray}
\Pi_{\rm static}^{(T=0)}&=& 4\kappa e^4
v^2\,\frac{\partial}{\partial m_\sigma^2}\int\frac{d^2k}{(2\pi)^2}\,
\vec{k}^2\, 
\left[\frac{1/\omega_\sigma}{(m_\sigma^2-m_+^2)(m_\sigma^2-m_-^2)}
-\frac{1/\omega_+}{(m_\sigma^2-m_+^2)(m_+^2-m_-^2)}
+\frac{1/\omega_-}{(m_\sigma^2-m_-^2)(m_+^2-m_-^2)}\right]
\end{eqnarray} 
These integrals may be performed with a consistent UV regulator, yielding
a finite result that agrees precisely with the zero temperature result
quoted in (\ref{ztpi}) (It is worth noting here that the Chern-Simons
coefficient in the Euclidean space is $i$ times that of the Minkowski space).

The finite temperature correction to this zero temperature result is
given by 
\begin{eqnarray}
\Pi_{\rm static}^{(\beta)}&=& 8\kappa e^4
v^2\,\frac{\partial}{\partial m_\sigma^2}\int\frac{d^2k}{(2\pi)^2}\,
\vec{k}^2\, 
\left[\frac{1}{(m_\sigma^2-m_+^2)(m_\sigma^2-m_-^2)}\frac{1}{\omega_{\sigma}}
\frac{1}{e^{\beta
\omega_\sigma}-1}
-\frac{1}{(m_\sigma^2-m_+^2)(m_+^2-m_-^2)}\frac{1}{\omega_{+}}\frac{1}
{e^{\beta\omega_+}-1}
\right.\nonumber\\
&&  \qquad\qquad\left.
+\frac{1}{(m_\sigma^2-m_-^2)(m_+^2-m_-^2)}\frac{1}{\omega_{-}}\frac{1}
{e^{\beta\omega_-}-1}
\right]
\label{ft}
\end{eqnarray}
Thus, we need to evaluate an integral of the form
\begin{eqnarray}
I=\int
\frac{d^2k}{(2\pi)^2}\,\frac{\vec{k}^2}{\omega}\,\frac{1}{e^{\beta\omega}-1}=
\frac{1}{2\pi\beta^3}\int_{m\beta}^\infty
dx\frac{\left(x^2-(m\beta)^2\right)}{e^x-1}
\label{integral}
\end{eqnarray}
where $\omega=\sqrt{\vec{k}^2+m^2}$. We note that
\begin{eqnarray}
\int_y^\infty dx\frac{1}{e^x-1}=-\log\left(1-e^{-y}\right)
\label{int1}
\end{eqnarray}
\begin{eqnarray}
\int_y^\infty dx\frac{x^2}{e^x-1}&=&\int_0^\infty
dx\frac{x^2}{e^x-1}-\int_0^y dx\frac{x^2}{e^x-1}\nonumber\\ 
&=&2\zeta(3)-\sum_{n=0}^\infty \frac{{\cal B}_n}{(n+2)\,n!}\, y^{n+2}
\label{int2}
\end{eqnarray}
where the ${\cal B}_n$ are the Bernoulli numbers.
Therefore, the high temperature expansion of $I$ is
\begin{eqnarray}
I&=&\frac{1}{2\pi
\beta^3}\left[(m\beta)^2\log(1-e^{-m\beta})+2\zeta(3)-\sum_{n=0}^\infty
\frac{{\cal B}_n}{(n+2)\,n!}(m\beta)^{n+2}\right]\nonumber\\
&= &\frac{1}{2\pi
\beta^3}\left[2\zeta(3)+(m\beta)^2\left(\log(m\beta)-\frac{1}{2}\right)+\dots\right]
\end{eqnarray}
Thus, in the static limit, at high temperature, the leading behavior is
\begin{eqnarray}
\Pi_{\rm static}^{(\beta)}= \frac{4\kappa e^4 v^2}{\pi\beta}
F(m_+,m_-,m_\sigma)
\label{static1}
\end{eqnarray}
where
\begin{eqnarray} F(m_+,m_-,m_{\sigma})&=&\frac{m_+^2
\log(m_+/m_-)}{(m_+^2-m_-^2)(m_{\sigma}^2-m_+^2)^2} + \frac{(m_+^2m_-^2-
m_{\sigma}^4) \log(m_{\sigma}/m_-)}{(m_{\sigma}^2 -m_+^2)^2
(m_{\sigma}^2-m_-^2)^2}+\frac{1}{2
(m_{\sigma}^2-m_+^2)(m_{\sigma}^2-m_-^2)}\nonumber\\ &=&\frac{m_-^2
\log(m_+/m_-)}{(m_+^2-m_-^2)(m_{\sigma}^2-m_-^2)^2} +
\frac{(m_+^2m_-^2-m_{\sigma}^4) \log(m_{\sigma}/m_+)}{(m_{\sigma}^2
-m_+^2)^2 (m_{\sigma}^2-m_-^2)^2}+ \frac{1}{2
(m_{\sigma}^2-m_+^2)(m_{\sigma}^2-m_-^2)}\nonumber\\ &=&\frac{m_+^2
\log(m_+/m_{\sigma})}{(m_+^2-m_-^2)(m_{\sigma}^2-m_+^2)^2} -\frac{m_-^2
\log(m_-/m_{\sigma})}{(m_+^2-m_-^2)(m_{\sigma}^2-m_-^2)^2}+
\frac{1}{2 (m_{\sigma}^2-m_+^2)(m_{\sigma}^2-m_-^2)}
\label{static2}
\end{eqnarray}
It is very interesting to notice that the temperature dependence inside the
logarithmic terms cancels out, leaving logarithms only involving ratios of the
masses $m_+$, $m_-$ and $m_\sigma$.

We now consider this result in the mass limits considered in Section III A.
First, we take the pure Chern-Simons limit in which $m_+\sim |\kappa|\to\infty$,
and $m_-\to \frac{m^2}{|\kappa|}={\rm finite}$. From the above, in this limit the
leading high temperature behavior is
\begin{eqnarray}
\Pi_{\rm static}^{(\beta)}\to - \frac{\kappa}{|\kappa|}\frac{e^2
m_-}{\pi\beta}\left[\frac{2m_-^2\log(\frac{m_-}{m_\sigma})+m_\sigma^2-m_-^2}
{(m_\sigma^2-m_-^2)^2}\right]
\end{eqnarray}
In the $N=2$ SUSY limit where we further require that the two remaining
masses are equal ({\it i.e.} $m_-=m_\sigma$), we find 
\begin{eqnarray}
\Pi_{\rm static}^{(\beta)}\to - \frac{\kappa}{|\kappa|}\frac{e^2}{2\pi\beta m_-}
\end{eqnarray}
Indeed, in this SUSY case we can keep the full temperature dependence.
Returning to the expressions (\ref{stat}) and (\ref{sw}) which have the
full temperature dependence, we can take the limit $m_+\to\infty$, and
$m_\sigma\to m_-$, to obtain a remarkable simplification:
\begin{eqnarray}
\Pi_{\rm static}\to - \frac{\kappa}{|\kappa|}\frac{e^2}{4\pi}\, {\rm
coth}\left(\frac{\beta m_-}{2}\right)
\label{bt}
\end{eqnarray}
At zero temperature this reduces to the result in (\ref{ztsusy}). As mentioned
before, at zero temperature, (\ref{ztsusy}) cancels against a fermion loop
contribution of equal magnitude but opposite sign. But at finite temperature,
the corresponding fermion loop contributes
\begin{eqnarray}
\Pi_{\rm static}= \frac{\kappa}{|\kappa|}\frac{e^2}{4\pi}{\rm
tanh}\left(\frac{\beta m_-}{2}\right)
\label{fft}
\end{eqnarray}
which does not cancel the bosonic loop contribution (\ref{bt}), except at zero
temperature. This is a reflection of the fact that the nonzero temperature
breaks the supersymmetry of the system \cite{daskaku}.

\subsubsection{Long Wavelength Limit}

In this Section we consider the long wavelength limit at finite temperature. In
this limit we set $\vec{p}=0$, and consider $p^0\to 0$. This must be done with
care in the imaginary time formalism, because $p^0$ is discrete and must be
analytically continued back to real time where it is a continuous variable.
In the long wavelength limit, the parity violating part of the polarization
tensor is
\begin{eqnarray}
\Pi^{{\rm LW}\,(PV)}_{\mu\nu}= 8\kappa e^4 v^2
\epsilon_{\mu\nu\lambda}\frac{1}{\beta}\sum_{n=-\infty}^\infty \int
\frac{d^2k}{(2\pi)^2}\, \frac{k_\lambda}{((k^0+p^0)^2+\vec{k}^2+m_\sigma^2)
((k^0)^2+\vec{k}^2+m_+^2)((k^0)^2+\vec{k}^2+m_-^2)}
\end{eqnarray}
By symmetry it is clear that
\begin{eqnarray}
\Pi^{{\rm LW}\,(PV)}_{0i}=0
\end{eqnarray}
while $\Pi^{{\rm LW}\,(PV)}_{ij}$ is nonzero. This is the opposite of what was found
above in the static limit, where $\Pi_{ij}^{{\rm static}\,(PV)} = 0$ and
$\Pi_{0i}^{{\rm static}\,(PV)} \neq 0$. In fact,
\begin{eqnarray}
\Pi^{{\rm LW}\,(PV)}_{ij}=  8 \kappa e^4 v^2
\epsilon^{ij}\frac{1}{\beta}\sum_{n=-\infty}^\infty \int
\frac{d^2k}{(2\pi)^2}\, \frac{k_0}{((k^0+p^0)^2+\vec{k}^2+m_\sigma^2)
((k^0)^2+\vec{k}^2+m_+^2)((k^0)^2+\vec{k}^2+m_-^2)}
\end{eqnarray}
The sum over Matsubara modes can be done, as before, using a contour integral
representation
\begin{eqnarray}
&&\frac{1}{\beta}\sum_{n=-\infty}^\infty 
\frac{k_0}{((k^0+p^0)^2+\vec{k}^2+m_\sigma^2)
((k^0)^2+\vec{k}^2+m_+^2)((k^0)^2+\vec{k}^2+m_-^2)}\nonumber\\
&&=-\frac{\pi}{\beta}\left(\frac{\beta}{2\pi}\right)^5 \sum_{\rm residues} \left[
\frac{z\, {\rm cot}(\pi z)}{\left[(z+\frac{\beta p_0}{2\pi})^2+(\frac{\beta
\omega_\sigma}{2\pi})^2\right]\left[(z+\frac{\beta p_0}{2\pi})^2+(\frac{\beta
\omega_+}{2\pi})^2\right]\left[(z+\frac{\beta p_0}{2\pi})^2+(\frac{\beta
\omega_-}{2\pi})^2\right]}\right]\nonumber\\
&&=\frac{1}{4}\left[\frac{{\rm coth}(\frac{\beta
\omega_\sigma}{2})}{\omega_\sigma}\,
\frac{(-p^0+i\omega_\sigma)}{[(-p^0+i\omega_\sigma)^2+\omega_+^2]
[(-p^0+i\omega_\sigma)^2+\omega_-^2]}-
\frac{{\rm coth}(\frac{\beta
\omega_\sigma}{2})}{\omega_\sigma}\,
\frac{(p^0+i\omega_\sigma)}{[(p^0+i\omega_\sigma)^2+\omega_+^2]
[(p^0+i\omega_\sigma)^2+\omega_-^2]}\right.\nonumber\\
&&\left. -\frac{i\,{\rm coth}(\frac{\beta
\omega_+}{2})}{[(p^0+i\omega_+)^2+\omega_\sigma^2]
[\omega_+^2-\omega_-^2]}+ \frac{i\,{\rm coth}(\frac{\beta
\omega_+}{2})}{[(p^0-i\omega_+)^2+\omega_\sigma^2]
[\omega_+^2-\omega_-^2]} - \frac{i\,{\rm coth}(\frac{\beta
\omega_-}{2})}{[(p^0+i\omega_-)^2+\omega_\sigma^2]
[\omega_+^2-\omega_-^2]} \right. \nonumber\\
&&\left. + \frac{i\,{\rm coth}(\frac{\beta
\omega_-}{2})}{[(p^0-i\omega_-)^2+\omega_\sigma^2]
[\omega_+^2-\omega_-^2]}\right]
\end{eqnarray}

This expression can now be analytically continued in $p^0$, and then Taylor
expanded to linear order in $p^0$, in order to determine the Chern-Simons
coefficient. We write
\begin{eqnarray}
\Pi^{{\rm LW}\,(PV)}_{ij} = \epsilon_{ij} p^0 \, \Pi_{\rm LW}(p^0)
\end{eqnarray}
where
\begin{eqnarray}
\Pi_{\rm LW}(p^0=0)&=& - 4\kappa e^4 v^2 \int
\frac{d^2k}{(2\pi)^2}\left[\left\{
\frac{1}{\omega_\sigma(m_\sigma^2-m_+^2)(m_\sigma^2-m_-^2)}-
\frac{2\omega_\sigma}{(m_\sigma^2-m_+^2)^2(m_\sigma^2-m_-^2)}
\right.\right.\nonumber\\
&&\left.\left.
-\frac{2\omega_\sigma}{(m_\sigma^2-m_+^2)(m_\sigma^2-m_-^2)^2}\right\}\, {\rm
coth}\left(\frac{\beta\omega_\sigma}{2}\right) +\frac{2\omega_+{\rm coth}
(\frac{\beta\omega_+}{2})}{(m_\sigma^2-m_+^2)^2(m_+^2-m_-^2)}\right.
\nonumber \\
&& \left. - \frac{2\omega_-{\rm coth}
(\frac{\beta\omega_-}{2})}{(m_\sigma^2-m_-^2)^2(m_+^2-m_-^2)}\right]
\label{lwfull}
\end{eqnarray}
Once again, we separate the zero temperature piece from the finite temperature
correction using the simple identity (\ref{identity}). Then the zero temperature
part of (\ref{lwfull}) is finite, with consistent UV regulators for the momentum
integrals, and agrees precisely with the direct zero temperature result in
(\ref{ztpi}).

The nonzero temperature contribution can be expressed as
\begin{eqnarray}
\Pi_{LW}^{(\beta)}&=& -\frac{4\kappa e^4
v^2}{\pi\beta^3}\left[\int_{\beta m_\sigma}^\infty
\frac{dx}{e^x-1}\left\{\frac{\beta^2}{(m_\sigma^2-m_+^2)(m_\sigma^2-m_-^2)}-
\frac{2x^2}{(m_\sigma^2-m_+^2)^2(m_\sigma^2-m_-^2)}-
\frac{2x^2}{(m_\sigma^2-m_+^2)(m_\sigma^2-m_-^2)^2}\right\}\right. \nonumber\\
&&\left. +\int_{\beta m_+}^\infty
\frac{dx}{e^x-1} \frac{2x^2}{(m_\sigma^2-m_+^2)^2(m_+^2-m_-^2)} -
\int_{\beta m_-}^\infty
\frac{dx}{e^x-1} \frac{2x^2}{(m_\sigma^2-m_-^2)^2(m_+^2-m_-^2)}\right]
\end{eqnarray}
The dominant contribution at high temperature is easily computed using the
integrals listed earlier in (\ref{int1},\ref{int2}). This dominant contribution
in the long wave limit, at high temperature, gives
\begin{eqnarray}
\Pi_{ij}^{{\rm LW}\, (PV)(\beta)} =  \frac{4  \kappa e^4 v^2
\epsilon^{ij}p_0}{\beta\pi} \,\frac{\log(\beta
m_{\sigma})}{(m_{\sigma}^2-m_+^2)(m_{\sigma}^2-m_-^2)}
\label{lw1}
\end{eqnarray}
We note several things about this result. First, there is still a logarithmic
dependence on the temperature. Second, the long wavelength limit gives a
completely different result for the parity violating part of the self energy, as
compared to the static limit. This is true even though the two masses in the
bosonic loop are quite different. So, there is still a non-analyticity in the
self-energy, contrary to what had been found earlier in a simpler model
\cite{arnold} without parity violation. For completeness, we note here
that, at high temperature, the contribution due to a fermion loop to
the Chern-Simons term goes as $\sim \beta$ in the static limit and as
$\sim \beta \ln \beta$ in the long wave limit \cite{brandt}.

\section{Conclusions}
To conclude, we emphasize that the induced Chern-Simons terms that appear from
bosonic loops have a completely different temperature dependence from those
induced by a fermion loop. For example, for the bosonic loops the static
limit of the graph grows at high temperature, while the fermionic loop
contribution vanishes at high temperature. This is most clearly seen by
comparing the static limit results (\ref{bt}) and (\ref{fft}) in the model with
mass parameters such that the zero temperature model has an N=2
supersymmetry. In fact, the contributions from the bosonic loop grows
at high temperature both in the static as well as the long wavelength
limits, as opposed to the contributions from the fermionic loop which
goes down with temperature in both these limits \cite{brandt}. 

In the model studied in this paper, the induced Chern-Simons terms arise in loop
corrections because of the presence of a parity violating bare Chern-Simons term
in the bare Lagrangian. This bare Chern-Simons term has a number of
consequences. First, it gives the gauge field two massive modes in the
spontaneously broken phase. Second, it introduces parity violating interactions.
We have shown that the induced parity violating contributions to the self energy
behave, at finite temperature, in a very different way from the parity
preserving contributions studied previously in \cite{arnold}. Specifically, the
parity preserving terms have a unique zero momentum limit, even at finite
temperature, while the parity violating terms have a non-unique limit at finite
temperature. We have demonstrated this by computing the parity violating terms in
both the static and long wavelength limit. The leading high temperature parts for
these parity violating terms are given in (\ref{static1},\ref{static2}) and
(\ref{lw1}), and they are clearly different. In the long wavelength limit there
is a logarithmic dependence on the temperature, while in the static limit this
cancels out leaving logarithmic dependence on mass ratios. 

In addition, we have analyzed the limit in which the mass parameters are such
that the zero temperature model has an N=2 supersymmetry, and have seen
explicitly how the SUSY is broken at finite temperature in the form of a
non-cancellation between bosonic loop and fermionic loop contributions. We
understand \cite{gomes} that M. Gomes and collaborators are analyzing a related
model involving a pure Chern-Simons gauge Lagrangian coupled to a Higgs field
with a sextic potential. It would be interesting to compare their results to
ours in the appropriate limit.

Finally, the temperature dependent parity violating contributions to the
self-energy mix with the temperature dependent parity conserving contributions
in order to determine the physical masses in thermal equilibrium. This issue has
been analyzed in \cite{pisarski1} for the Chern-Simons-Yang-Mills system at
finite temperature. A similar analysis for the Maxwell-Chern-Simons-Higgs
system with symmetry breaking would be an interesting application of the
results in this current paper.

\bigskip
\noindent{\bf Acknowledgement:} GD thanks the DOE for support through grant
DE-FG02-92ER40716.00, and AD thanks the DOE for support through grant
DE-FG-02-91ER40685. VSA and SP are supported through CAPES, Brasil.

\end{document}